\documentclass[twocolumn,twoside,slac_two]{revtex4}
\usepackage{graphicx}
\usepackage{fancyhdr}
\begin{document}
   
   \title{Rare {\boldmath $B$} decays and new physics studies}

   %
   
   \author{Owen Long}
   \affiliation{University of California at Riverside, Riverside CA 92521, USA}

   \begin{abstract}
      I present a review of using rare $B$ decays
      to search for physics beyond the Standard Model.
      $B$ decays that proceed either through annihilation
      or loop topologies at leading order in the
      Standard Model provide unique probes in the
      search for new physics.  The latest experimental
      results from the $B$ factories (Babar and Belle)
      and the Tevatron experiments (CDF and D0) on
      rare decays and their impact on various scenarios
      for new physics will be presented.
   \end{abstract}
   
   \maketitle
   
   \thispagestyle{fancy}

   \section{Introduction}

   In the past decade, we have seen enormous progress in
   understanding flavor physics and $CP$ violation.
   After turning on in 1999, the new asymmetric-energy $B$
   factories, PEP-II~\cite{pepref} and KEKB~\cite{kekbref},
   quickly achieved luminosities that exceeded their design
   targets and the expectations of many.
   This allowed the corresponding experiments, Babar~\cite{babarref}
   and Belle~\cite{belleref}, to quickly provide the first
   precision test of the CKM~\cite{ckm} mechanism for $CP$ violation.
   The measurements of the proper-time-dependent $CP$ asymmetry in
   charmonium-$K^0$ decays of neutral $B$ mesons ($\sin2\beta$)\cite{sin2beta-alpha}
   are in very good agreement with the CKM prediction of $\sin2\beta$
   from independent constraints.

\begin{figure}
     \includegraphics[width=75mm]{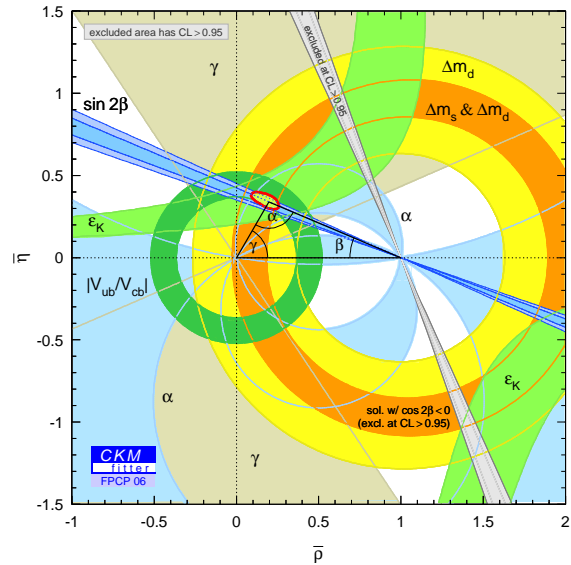}
     \caption{Constraints on the Unitarity Triangle as of
       the FPCP 2006 conference.  Figure courtesy of the
       CKMfitter group\cite{ckmfitter}.} \label{fig:rhoeta}
     \end{figure}

   It is convenient, both for visualization and quantitative
   analysis, to interpret experimental results within the CKM
   framework as constraints on the geometry of the so-called "Unitarity
   Triangle"~\cite{kirkby-nir}, which is from the first and third columns of
   the CKM quark mixing matrix $V_{ij}$
   \begin{equation}
      V_{ub}^*V_{ud} + V_{cb}^*V_{cd} + V_{tb}^*V_{td} = 0 .
   \end{equation}
   If one renormalizes the triangle by rescaling the
   sides by $1/V_{cb}^*V_{cd}$ and adopts the Wolfenstein
   phase convention~\cite{wolfenstein}, experimental results
   are interpreted as constraints on the apex of the triangle
   ($\bar \rho, \bar \eta$).
   Two independent groups (CKMfitter\cite{ckmfitter}
   and UTfit\cite{utfit}) provide the results of this analysis.
   Figure~\ref{fig:rhoeta}  shows the constraints on the apex of the Unitarity
   Triangle as of the FPCP'06 conference.
   One can see that in addition to the precise determination of
   $\beta$, mentioned above, the $B$ factory data strongly
   constrains the left side of the triangle, which is
   proportional to $|V_{ub}|$\cite{vub-gamma}.
   The $B$ factory experiments have also measured the other
   two angles of triangle ($\alpha$\cite{sin2beta-alpha}
   and $\gamma$\cite{vub-gamma}).
   Finally, the Tevatron experiments, CDF and D0, have recently
   measured $\Delta m_s$\cite{deltams}, the frequency of $B_s$ oscillations,
   which allows the right side of the triangle to be constrained,
   when combined with $\Delta m_d$ and some input from
   lattice QCD calculations\cite{lattice}.

   It's clear from Figure~\ref{fig:rhoeta} that the CKM model (and thus the
   Standard Model) gives a remarkably consistent description of
   all experimental results.
   It's natural to ask whether there is still room for discovering
   physics beyond the Standard Model (or New Physics) in $B$ decays,
   after seeing such consistency.
   The answer is yes.
   The most precise constraints in Figure~\ref{fig:rhoeta} come from either
   tree-level $B$ decays or from $B$ mixing.
   From Figure~\ref{fig:rhoeta}, one can reasonably conclude that there is
   little room for substantial New Physics corrections to
   tree-dominated processes~\cite{silvestrini}.
   However, New Physics may significantly alter the
   observables (branching fractions, asymmetries, ...)
   for rare $B$ decays without disturbing the beautiful
   consistency shown in Figure~\ref{fig:rhoeta}.

     Rare $B$ decays are a unique and valuable tool in the search
     for New Physics.
     Decays that are allowed at the tree level have relatively large
     amplitudes.
     If amplitudes from New Physics are small, as we expect,
     hadronic uncertainties will prevent us from recognizing the
     presence of New Physics in many cases.
     On the other hand, decays that proceed through annihilation
     or loop topologies at leading order are highly suppressed, thus
     considerably reducing the impact of hadronic uncertainties in the
     search for New Physics.
     Loop topologies, such as penguin decays, are particularly attractive.
     Virtual new particles (e.g. supersymmetric particles) with masses
     on the order of 100's of GeV may contribute loop contributions
     to these decays potentially altering the decay rate, $CP$
     asymmetry, and other observable quantities.

   It is important to have Standard Model predictions for rare decay
   observables with theoretical uncertainties that are under control
   and as low as possible.
   An excellent way to do this is to have uncertainties cancel in a ratio,
   such as a $CP$ asymmetry.
   Time-dependent $CP$ asymmetries in penguin-dominated decays, such
   as $\phi K^0$ and $\eta'K^0$ are a prime example that I will discuss.
   Another way is to avoid hadrons in the final state.
   I will review the latest results on $B$ decays to $\mu \mu$,
   $\tau \nu$, $s \gamma$, $d \gamma$, and $s \ell \ell$ final states.
   Finally, one can use both techniques.
   For example, the $CP$ asymmetry in $b\rightarrow s\gamma$ or the
   forward-backward asymmetry in $B\rightarrow K^* \ell \ell$.

     The list of topics above is certainly not a complete inventory
   of the avenues being pursued.
   I chose to focus on them because they are areas where recent
   progress has been made and/or because they are channels with
   relatively small theoretical uncertainties.
   Before continuing, I would like to point out that there have been
   very good recent reviews on many of the topics that I will cover,
   such as reference~\cite{schetinger}, that the reader may be
   interested in for more details or a different point of view.


   \section{\boldmath $CP$ asymmetries in $b\rightarrow s$ penguin decays}

     Time-dependent $CP$ asymmetries in $b\rightarrow s$ penguin decays
     provide an excellent way to search for new physics~\cite{btos-np}.
     As I will describe below, most hadronic uncertainties cancel
     in the Standard Model calculation of the expected asymmetry,
     so there are precise predictions to compare our measurements with.
     Contributions from non-Standard-Mode particles may give
     large (order 1) corrections to the $CP$ asymmetries.
     A large deviation from the Standard Model expectation would
     be an unambiguous sign of New Physics.

     The Standard Model interpretation of a time-dependent $CP$
     asymmetry is theoretically clean {\em if} one decay amplitude
     dominates (or, more technically, if all dominant decay amplitude
     contributions share the same $CP$-violating phase).
     The most familiar example of this is $B^0\rightarrow J/\psi K^0_S$.
     The leading decay amplitude is a color-suppressed tree diagram.
     The largest amplitude with a different weak phase comes from
     a $b\rightarrow s$ penguin diagram with an intermediate $u$ quark,
     which is suppressed with respect to the dominant amplitude by
     a relative $CKM$ factor of about 0.02, loop vs tree suppression,
     and by the need to create a $c\bar c$ pair.
     Thus, the Standard Model predicts that the amplitude of the
     time-dependent $CP$ asymmetry for the $J/\psi K^0_S$
     final state is $\sin 2\beta$ with an uncertainty of less than
     1\%~\cite{jpsiks-clean}.

     The time-dependent $CP$ asymmetry for decays that proceed
     through a penguin $b\rightarrow s$ transition is also
     expected to be close to $\sin2\beta$.
     However, the corrections from suppressed $b\rightarrow u$
     amplitudes are in general larger.
     For $b\rightarrow s\bar s s$ transitions, such as
     $B^0 \rightarrow \phi K^0$ or
     $B^0 \rightarrow K^0_S K^0_S K^0_S$,
     the only $b\rightarrow u$ amplitude comes from the $b\rightarrow s$
     penguin diagram with an intermediate $u$ quark.
     Here, both the leading and CKM suppressed contributions are
     both penguins, differing only by the intermediate quarks
     in the loop.
     For decays such as $B^0\rightarrow \eta' K^0$, a color-suppressed
     $b\rightarrow u$ tree diagram may also contribute, since the
     $\eta'$ has both $u\bar u$ and $s\bar s$ content.
     In this situation, the dominant amplitudes are loop suppressed,
     so the $b\rightarrow u$ tree contribution is more of a concern.
     The key theoretical task is to determine or place an upper
     bound on the relative size of the $b\rightarrow u$ amplitudes,
     which involves calculating or constraining the ratio of
     hadronic matrix elements.

     There has been a tremendous amount of theoretical work on
     estimating the Standard Model uncertainty on the $CP$ asymmetry
     in $b\rightarrow s$ penguin decays~\cite{btos-theory}.
     Theoretical estimates are given in terms of the maximum deviation
     from $\sin2\beta$ of
     the measured $\sin \Delta m \Delta t$ coefficient $S_f$ of
     the $CP$ asymmetry for final state $f$,
     or $\Delta S_f = S_f - \sin2\beta$.
     Conservative model-independent estimates of $|\Delta S_f|$
     are on the same order or larger than the current experimental
     errors.
     Model dependent calculations estimate $\Delta S_f$, including
     its sign, and have uncertainties at the level of 1 to 2 \%.

     The most interesting channels, both experimentally and theoretically,
     are $\phi K_0$ and $\eta' K_0$.
     The $\phi K_0$ mode is a pure $b\rightarrow s\bar s s$ transition
     with a clean experimental signature.
     The $\eta' K_0$ mode is not a pure $b\rightarrow s\bar s s$ transition
     but model estimates predict a very small $\Delta S_f$~\cite{btos-qcdf}.
     The $\eta' K_0$ mode also has an unusually large branching fraction
     so the statistical errors on the $S_f$ measurements
     are best for this decay.
     As with the theoretical uncertainties, many experimental
     systematic uncertainties cancel in the $CP$ asymmetry.
     Systematic uncertainties on the $S_f$ measurements are 3 to 5
     times smaller than the statistical errors.
     Both the Babar and Belle experiments reconstruct the $K^0_S$ and
     the more challenging $K^0_L$ final states in order to maximize
     their signal statistics.

     \begin{figure}
     \includegraphics[width=75mm]{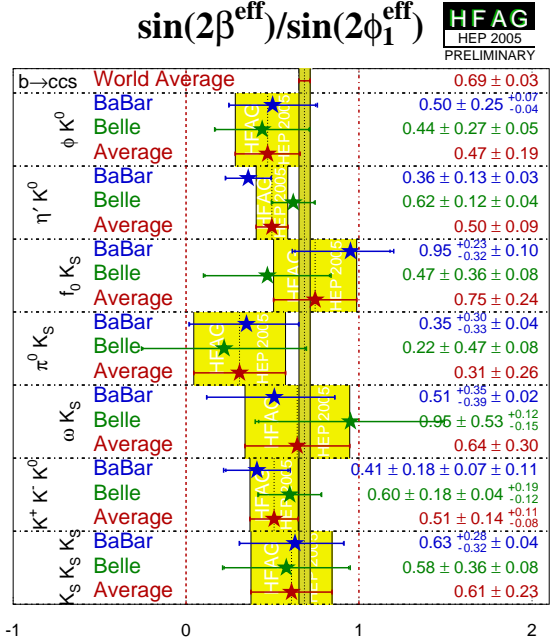}
     \caption{ Measurements of $S_f$ for several $b\rightarrow s$ penguin
       decays.  The Standard Model predicts 
       $-\eta_f \, S_f = S_{b\rightarrow c\bar c s} = \sin 2 \beta$ if $b\rightarrow u$
       amplitude contributions are neglected.
       Figure courtesy of the
       Heavy Flavor Averaging Group (HFAG)\cite{hfag}.} \label{fig:speng}
     \end{figure}

     Figure~\ref{fig:speng} summarizes the $S_f$ measurements for
     many $b\rightarrow s$ penguin decay modes, as compiled by
     the Heavy Flavor Averaging Group (HFAG)\cite{hfag}.
     None of the channels show a significant discrepancy with the
     Standard Model expectation of
     $-\eta_f \, S_f = S_{b\rightarrow c\bar c s} \approx \sin 2\beta$,
     where $\eta_f$ is the $CP$ eigenvalue of the final state $f$.
     The measurements tend to be below $\sin2\beta$ (i.e. $\Delta S_f<0$)
     which is interesting because model predictions give
     small {\em positive} values of $\Delta S_f$~\cite{btos-qcdf}.
     There is still room for discovery, since the measurements are
     and will be statistics limited for the foreseeable future,
     though a super $B$ factory~\cite{superB} may be required to achieve the
     sensitivity necessary for a discovery.


    \section{ \boldmath Leptonic $B$ decays }

    $B$ decays to purely leptonic final states are theoretically
    advantageous, since the hadronic uncertainties mainly come
    from a single parameter, the $B_i$ meson decay constant $f_i$.
    The $B$ decay constant can be computed using lattice
    QCD~\cite{lattice} or measured from a leptonic $B$ decay
    branching fraction, as we will see.
    New physics can greatly enhance leptonic $B$ decay rates in
    some scenarios, so searching for them is a good way to search
    for New Physics.


    \subsection{ \boldmath The search for $B_s \rightarrow \mu^+ \mu^-$}

    The decay $B_s \rightarrow \mu^+ \mu^-$ is highly suppressed
    in the Standard Model, since it can only proceed through
    a box diagram or a $Z$ penguin.
    The current SM prediction is
    ${\cal B}(B_s \rightarrow \mu^+ \mu^-) = (3.4 \pm 0.5) \times 10^{-9}$~\cite{btomumu-pred}.
    In some new physics scenarios, the branching fraction can
    be enhanced by by a high power of $\tan \beta$ (e.g.
    ${\cal B} \propto \tan^6 \beta$ in supersymmetry or
    ${\cal B} \propto \tan^4 \beta$ in 2 Higgs doublet models),
    where $\tan \beta$ is the ratio of the Higgs vacuum expectation values.
    For large $\tan \beta$, the branching fraction could be enhanced
    by two orders of magnitude, which is currently within reach of the CDF
    and D0 experiments.

    The Tevatron has a 4-order-of-magnitude advantage over the
    $B$ factories (Babar and Belle) in the $b\bar b$ production
    cross section.
    It also produces all types of $b$-quark hadrons, while
    $B_s \overline{B_s}$
    is above threshold for the $\Upsilon(4S)$ resonance,
    where most of the $B$ factory data have been collected.
    The main disadvantage of the Tevatron is that the $b\bar b$
    cross section is only $\approx 1/1000$ of the total inelastic
    $p \bar p$ cross section, so for most $B$ decays, QCD background
    is a major obstacle.
    Fortunately, both CDF and D0 can easily trigger on muons.

    The CDF experiment has recently updated their search for
    $B_s \rightarrow \mu^+ \mu^-$ with 780~pb$^{-1}$ of
    Run II data~\cite{CDFbstomumu}.
    They use a likelihood ratio that combines three discriminating
    variables to reject combinatoric background for events
    with a di-muon invariant mass close to the $B_s$.
    The variables are the reconstructed decay length,
    the consistency between the the vector that points from the
    primary vertex to the di-muon vertex and the di-muon momentum
    vector, and the degree to which the muons are isolated in
    the event.
    After requiring the likelihood ratio to be greater than
    0.99, the expected background is 1.3 events.
    They observed one event in the signal box, consistent with
    the background estimate, and set an upper limit of
    ${\cal B}(B_s \rightarrow \mu^+ \mu^-) < 1.0 \times 10^{-7}$
    at 95\% confidence level.
    This is only a factor of 30 above the Standard Model prediction.
    The D0 collaboration last updated their $B_s \rightarrow \mu^+ \mu^-$
    search using 300~pb$^{-1}$ of Run II data~\cite{D0bstomumu}.

    It has been noted that ${\cal B}(B_s \rightarrow \mu^+ \mu^-)$
    has connections to cosmology and dark matter.
    For example, reference~\cite{bstomumu-np} has pointed out that
    the $B_s \rightarrow \mu^+ \mu^-$ branching fraction is strongly
    correlated with the neutralino-proton scattering cross section
    for large $\tan \beta$.
    The lack of an observation of this mode at the Tevatron
    disfavors a large $\tan \beta$ and implies that the
    neutralino-proton cross section is well below the sensitivity
    of direct detection experiments, such as CDMS~II.


    \subsection{ \boldmath Evidence for $B^+\rightarrow \tau^+ \nu$ }

    The leptonic decay $B^+\rightarrow \tau^+ \nu$ is the least
    helicity suppressed, as can be seen by the relation below for
    the branching fraction:
    \begin{equation}
      {\cal B}(B^+\rightarrow \tau^+ \nu)
       = \frac{G^2_F \, m_B \, m_\tau^2}{8\pi} \,
          \left( 1 - \frac{m_\tau^2}{m_B^2} \right)
           \ f_B^2 \, \left| V_{ub} \right|^2 \tau_B .
    \end{equation}
    The Standard Model prediction for the branching fraction
    is $(1.50 \pm 0.40)\times 10^{-4}$~\cite{Belle-btotaunu},
    where the uncertainty
    is dominated by the calculation of $f_B$ from lattice
    QCD ($\approx 10\%$)\cite{fb-lattice} and $|V_{ub}|$
    from semileptonic $B$ decays ($\approx 7.5\%$)\cite{hfag}.
    This is now within reach of the current $B$ factories.
    Measuring this branching fraction is the most straightforward
    way to experimentally test the lattice calculation of
    the $B$ decay constant $f_B$.
    Large enhancements of the branching fraction are possible
    in some New Physics models, such as the two-Higgs-doublet
    model described in reference~\cite{btaunu-2higgs-doublet}
    with large $\tan \beta$.
    The Belle experiment has recently
    reported evidence~\cite{Belle-btotaunu} for
    $B^+\rightarrow \tau^+ \nu$.
    I will describe their analysis in the remainder of this
    subsection.

    This decay is very challenging due to the presence of at least
    two neutrinos in the final state.
    In $B$ decays with only one neutrino, the neutrino can be
    reconstructed from the missing momentum of the event, assuming
    the neutrino is massless.
    With two or more neutrinos, the missing invariant mass is,
    in general, non-zero, so the kinematics can't be inferred
    from the missing momentum alone.
    The technique that has been used by both the Belle and Babar
    experiments has been to fully reconstruct the other $B$
    (or the ``tag'' $B$) in the event and then search for the
    $B^+\rightarrow \tau^+ \nu$ decay in the recoil.
    The tag $B$ reconstruction efficiency is around $0.14$\%.

    Fully reconstructing the tag $B$ serves two purposes.
    First, it determines the {\it a priori} unknown
    direction of the signal $B$ momentum vector.
    Secondly, and more importantly, it allows a full reconstruction
    of the event (minus the neutrinos).
    The signal $\tau$ decay is reconstructed
    with an efficiency of 33\% in 5 decay modes
    that correspond to 81\% of all $\tau$ decays:
    $e^- \bar \nu_e \nu_\tau$,
    $\mu^- \bar \nu_\mu \nu_\tau$,
    $\pi^- \nu_\tau$,
    $\pi^- \pi^0 \nu_\tau$,
    and $\pi^- \pi^+ \pi^- \nu_\tau$.
    Since all of the observable decay products (everything
    but the neutrinos) have been accounted for in the reconstruction
    of the tag $B$ and the $\tau$,
    there should be no additional tracks in the event
    and the unaccounted-for calorimeter energy should be close to zero.

    The main discriminating variable in the Belle analysis,
    after all selection criteria are applied, is the remaining
    (extra) energy in the electromagnetic calorimeter $E_{ECL}$, which
    should be consistent with zero for signal events.
    The number of expected background events in a defined
    signal region near zero is $32.8 \pm 4.6$.
    A total of 54 events were observed in the signal region
    indicating the presence of a signal.
    An unbinned likelihood fit of the $E_{ECL}$ spectrum gave
    a fitted signal yield of $21.2^{+6.7}_{-5.7}$ events for
    a significance of $4.2~\sigma$, including systematic uncertainties.
    This corresponds to a $B^+\rightarrow \tau^+ \nu$ branching
    fraction~\footnote{
    An error was found in the analysis documented in~\cite{Belle-btotaunu}.
    This was reported at the ICHEP'06 conference in Moscow.
    The corrected branching fraction is
      $(1.79 \ ^{+0.56}_{-0.49} \ ^{+0.39}_{-0.46}) \times 10^{-4}$
    with a signal significance of $3.5~\sigma$, including systematic errors.
    As of the writing of the report, reference~\cite{Belle-btotaunu}
    has not been updated to correct the reported error.
    }
    of $(1.06 \ ^{+0.34}_{-0.28} \ ^{+0.18}_{-0.16}) \times 10^{-4}$,
    which is consistent with the Standard Model expectation of
    about $1.5 \times 10^{-4}$.

    The fact that an excess over the Standard Model prediction was
    not observed rules out a region of charged Higgs mass vs $\tan \beta$
    in the model described in~\cite{btaunu-2higgs-doublet}.
    In the future, improved measurements of
    ${\cal B}(B^+\rightarrow \tau^+ \nu)$ will provide a valuable
    experimental benchmark for the lattice QCD prediction of $f_B$.

    
    \section{ \boldmath Search for New Physics in $b\rightarrow s \, \gamma$
          and $b\rightarrow s \, \ell^+ \ell^-$ }

    The phenomenology of $b\rightarrow s \, \gamma$ and
    $b\rightarrow s \, \ell^+ \ell^-$ decays is closely linked.
    Standard Model calculations for these rare decays
    are performed using an effective Hamiltonian that is
    written in terms of several short-distance operators~\cite{ali-bsg-bsll-theory}.
    Wilson coefficients quantify the relative strength
    of the different short-distance contributions.
    Standard Model predictions for observable quantities,
    such as differential decay rates, can be written in terms
    of effective Wilson coefficients, which include higher-order
    corrections.
    The process $b\rightarrow s \, \gamma$ is dominated by the
    photon penguin operator, with Wilson coefficient
    $C_7$, while $b\rightarrow s \, \ell^+ \ell^-$ also has contributions
    from semileptonic vector and axial-vector operators with
    Wilson coefficients $C_9$ and $C_{10}$ respectively.

    Experimentally, we wish to measure $C_7$, $C_9$, and $C_{10}$.
    New Physics may alter the magnitude and/or sign of the
    effective Wilson coefficients.
    If a deviation from the Standard Model is observed, it can
    be used to distinguish between different models for New Physics,
    since the New Physics predictions for the deviations are not
    universal~\cite{hiller}.


    \subsection{ \boldmath Status of $b\rightarrow s \, \gamma$ }

    Gino Isidori recently called $b \rightarrow s \, \gamma$ ``The
    most effective NP killer''~\cite{isidori-quote}.
    There are two complimentary experimental approaches.
    In the semi-inclusive $B\rightarrow X_s \, \gamma$ approach, several 
    $X_s$ states are explicitly reconstructed (e.g. $K\pi$, $K\pi\pi$,
    $KKK\pi$, etc...).
    The total $B\rightarrow X_s \, \gamma$ rate is then computed
    by estimating the contribution from the missing $X_s$ states.
    The uncertainty in estimating the contribution from missing states
    is significant.
    On the other hand, only the photon is explicitly reconstructed
    in the fully inclusive approach.
    The huge background from continuum events ($e^+e^- \rightarrow q\bar q$
    with $q=u,d,s,$ or $c$)
    is suppressed by requiring the presence of a high-momentum lepton
    from the semileptonic decay of the other $B$ in the event.
    The remaining continuum background is then subtracted using
    $e^+e^-$ data taken just below the $\Upsilon(4S)$ resonance.
    The two methods have similar precision.

    The current HFAG world average~\cite{hfag} for $E_\gamma > 1.6$~GeV is
    \begin{equation}
     {\cal B}(b\rightarrow s \, \gamma)_{\rm expt.} =
       \left( 355 \pm 24 \ ^{+9}_{-10} \pm 3\right) \times 10^{-6} ,
    \end{equation}
    where the uncertainties, from left to right, are from experimental
    statistical and systematic sources,
    the $E_\gamma$ shape function, and the subtraction of
    $b\rightarrow d \, \gamma$.
    The next-to-leading-order theoretical prediction of the
    same quantity~\cite{btosgamma-pred} is
    \begin{equation}
     {\cal B}(b\rightarrow s \, \gamma)_{\rm theory} =
       \left( 357 \pm 30 \right) \times 10^{-6}.
    \end{equation}
    The remarkable agreement effectively fixes $|C_7|$ to the
    Standard Model value.
    However, the inclusive $b\rightarrow s \gamma$ rate is
    not sensitive to the {\em sign} of $C_7$.

    New physics can hide within $b\rightarrow s \, \gamma$
    without affecting the overall rate significantly.
    The $CP$ asymmetry
    \begin{equation}
      A_{CP}(b\rightarrow s \, \gamma)
             = \frac{ \Gamma(b \rightarrow s\, \gamma)
                    - \Gamma(b \rightarrow s\, \gamma) }
                    { \Gamma(b \rightarrow s\, \gamma)
                    + \Gamma(b \rightarrow s\, \gamma) }
    \end{equation}
    is extremely small within the Standard Model~\cite{hurth-btosgamma-acp}
    and is predicted to be
    $(0.42 ^{+0.17}_{-0.12})\%$.
    A large $A_{CP}$ would be a smoking gun for New Physics and
    would tell us something about the nature of the New Physics.
    In some scenarios, such as minimal flavor violation~\cite{mfv},
    $A_{CP}$ remains below 2\%, while others allow $A_{CP}$ to
    be of order 10\%.
    The HFAG average~\cite{hfag} experimental value is
    \begin{equation}
      A_{CP}(b\rightarrow s \, \gamma)_{\rm expt.} =
        \left( 0.4 \pm 3.6\right)\% ,
    \end{equation}
    which is consistent with the Standard Model prediction.
    The statistical errors on the $B$ factory
    measurements~\cite{babar-btosgamma-acp}\cite{belle-btosgamma-acp}
    are about 5\% while the systematic errors are 2.6\% and 1.5\%
    respectively.
    These measurements are based on 80~fb$^{-1}$ and 140~fb$^{-1}$
    of on-resonance data, so the precision of the world average
    will continue to improve.


    \subsection{ \boldmath Observation of $b\rightarrow d \, \gamma$ }

    The amplitude for the process
    $b\rightarrow d \, \gamma$ is suppressed by a factor
    of $|V_{td}/V_{ts}|\approx 0.2$
    with respect to $b\rightarrow s \, \gamma$.
    Smaller amplitudes, in general, have higher sensitivity to
    New Physics contributions.
    The branching fraction is suppressed by a factor of $\approx 0.04$
    making this channel extremely challenging.
    If the $b\rightarrow d \, \gamma$ rate is measured, it can
    be combined with the $b\rightarrow s \, \gamma$ rate to constrain
    $|V_{td}/V_{ts}|$, which is essentially the right side of
    the Unitarity Triangle, since many of the hadronic uncertainties
    cancel in the ratio of exclusive modes.

    Belle has recently observed the $b\rightarrow d \, \gamma$
    process for the first time~\cite{belle-btodgamma}, by
    exclusively reconstructing $B\rightarrow \rho \gamma$ and
    $B\rightarrow \omega \gamma$.
    They combined their exclusive measurements
    using the following isospin relation
    \begin{eqnarray*}
       \Gamma(B\rightarrow (\rho,\omega)\gamma) & \equiv &
       \Gamma(B^- \rightarrow \rho^- \, \gamma) \\
      & = & 2 \times \Gamma(\overline{B^0} \rightarrow \rho^0 \, \gamma) \\
      & = & 2 \times \Gamma(B \rightarrow \omega \, \gamma)
    \end{eqnarray*}
    and found
    ${\cal B}(B\rightarrow (\rho,\omega) \gamma )
      = \left( 1.32 \ ^{+0.34}_{-0.31} \ ^{+0.10}_{-0.09} \right)
        \times 10^{-6}$.
    which is consistent with the Standard Model expectation.~\footnote{
    The Babar experiment reported a confirmation of the Belle
    $b\rightarrow d \, \gamma$ observation at the ICHEP'06 conference.}


    \subsection{ \boldmath Recent progress in 
    $b\rightarrow d \, \ell^+ \ell^-$ }

    The process $b\rightarrow d \, \ell^+ \ell^-$ has contributions
    from the 3 short-distance operators that correspond to the
    Wilson coefficients $C_7$, $C_9$, and $C_{10}$.
    The relative strength of the contributions depends on
    $q^2$, which is the square of the di-lepton invariant mass.
    Differential measurements as a function of $q^2$ measure
    the amount of $A-V$ interference, thus constraining the
    Wilson coefficients.

    Some New Physics models wiggle out of the stringent constraint
    on $|C_7|$ from $b\rightarrow s \, \gamma$ by simply changing
    the {\em sign} of $C_7$.
    In $b\rightarrow d \, \ell^+ \ell^-$, changing the sign of
    $C_7$ would enhance the overall $b\rightarrow d \, \ell^+ \ell^-$
    rate, particularly in the low-$q^2$ range, essentially by
    changing destructive interference into
    constructive interference~\cite{wrong-sign-c7-theory-goto}.

    The inclusive $b\rightarrow d \, \ell^+ \ell^-$ rate has been
    measured in bins of $q^2$ by both Babar and Belle.
    They use the sum of many exclusive modes composed of
    one $K_S$ or $K^\pm$ plus zero or more pions.
    The rate from $K_L$ modes is estimated from the measured $K_S$ rates.
    The regions of $q^2$ near the charmonium resonances are excluded
    from the measurement and provide a very useful control sample.
    The sum of exclusive modes is extrapolated to find the fully
    inclusive rate.

\begin{figure}[h]
     \includegraphics[width=85mm]{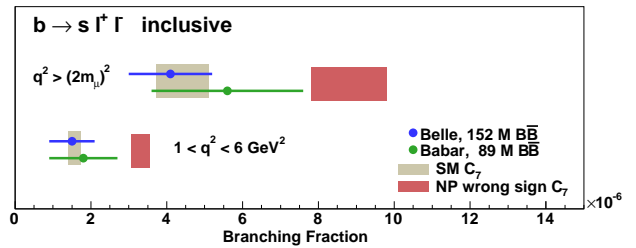}
     \caption{ Comparison of the Belle~\cite{belle-btosll-inclusive}
     and Babar~\cite{babar-btosll-inclusive} decay rates with
     theoretical predictions~\cite{btosll-inclusive-theory}
     for the Standard Model sign of $C_7$ and the opposite
     (or wrong) sign of $C_7$.  The data clearly favor the
     Standard Model sign of $C_7$.}
     \label{fig:btosll-rate-vs-q2}
     \end{figure}

    Figure~\ref{fig:btosll-rate-vs-q2} shows a comparison of
    the Belle~\cite{belle-btosll-inclusive} and
    Babar~\cite{babar-btosll-inclusive} measurements with
    theoretical predictions~\cite{btosll-inclusive-theory}
    for the Standard Model sign of $C_7$ and the opposite
    (or wrong) sign of $C_7$.
    The data clearly favor the Standard Model sign of $C_7$.

    A very interesting observable in $B\rightarrow K^* \ell^+ \ell^-$
    decays is the forward-backward asymmetry of the di-lepton system,
    which is defined as
    \begin{equation}
      A_{FB}(q^2) =
      \frac{   \Gamma(q^2, \cos\theta_{B\ell^-} > 0)
             - \Gamma(q^2, \cos\theta_{B\ell^-} < 0) }
           {   \Gamma(q^2, \cos\theta_{B\ell^-} > 0)
             + \Gamma(q^2, \cos\theta_{B\ell^-} < 0) },
    \end{equation}
    where $\cos\theta_{B\ell^-}$ is the angle between the
    negatively (positively) charged lepton direction and the
    $B$ ($\overline{B}$) flight direction in the di-lepton
    center-of-mass frame.
    Features of the $A_{FB}(q^2)$ curve can test both the
    signs and magnitudes of $C_9$ and $C_{10}$~\cite{ali-bsg-bsll-theory}.

    In the Standard Model, $A_{FB}(q^2)$ is predicted to
    be zero at $q^2=0$, dip negative, cross zero at about
    $q^2 = 0.2 \, m_b^2$, and rise monotonically to about
    0.5 at $q^2=1$.
    New Physics can drastically change the shape of the
    $A_{FB}(q^2)$ curve.
    For example, sign of $A_{FB}(q^2)$ can be flipped,
    the zero-crossing point may be shifted, or $A_{FB}(q^2)$
    may not even cross zero~\cite{ali-bsg-bsll-theory}.

    Both Babar and Belle have established signals in the
    $K^* \ell^+ \ell^-$ channels,
    though the statistics are low:
    about signal 50 events for Babar~\cite{babar-kstarll}
    and 100 for Belle~\cite{belle-kstarll}.
    Backgrounds are also non-negligible.
    The background fraction in the signal region is roughly 50\%.
    This is not enough for a precision scan of $A_{FB}$ vs
    $q^2$, but the experiments have shown that interesting
    conclusions can already be made from the currently
    analyzed data.

    Both experiments favor a positive integrated asymmetry,
    as expected in the Standard Model.
    The results are given in Table~\ref{tab:kll-afb} for both
    $K^*\ell\ell$ and for $K\ell\ell$ where no asymmetry is
    expected.
    The Belle analysis measured $A_{FB}$ in five bins of $q^2$
    and then fit for the leading relative terms $A_9/A_7$ and $A_{10}/A_7$
    of the effective Wilson coefficients.
    The fit is very consistent with the Standard Model prediction
    and is able to exclude a wrong-sign $A_9 A_{10}$ combination
    at 98.2\% confidence level.

    \begin{table}
      \begin{tabular}{|l|c|c|}
      \hline
          &  $\int A_{FB}(K^*\ell\ell)$  &  $\int A_{FB}(K\ell\ell)$ \\
      \hline
        Belle~\cite{belle-kstarll} &  $0.50 \pm 0.15 \pm 0.02 \ (3.4\sigma)$
            &  $0.10 \pm 0.14 \pm 0.01$ \\
        Babar~\cite{babar-kstarll} &  $>0.55$ @ 95\% C. L.
            &  $0.15 ^{+0.21}_{-0.23}\pm 0.08$ \\
      \hline
      \end{tabular}
      \caption{Di-lepton forward-backward asymmetries integrated
         over $q^2$ ($\int A_{FB}$) for $B\rightarrow K^{(*)}\ell\ell$.
         A positive net asymmetry is expected for $K^*\ell\ell$
         in the Standard Model, while no asymmetry is expected
         for $K\ell\ell$.
         }
      \label{tab:kll-afb}
    \end{table}

    The ultimate reach of the $B$ factories will still probably
    leave these measurements limited by statistical uncertainties.
    LHCb will pick up where the current $B$ factories will leave
    off and hopefully realize the full discovery potential
    of the $K^*\ell\ell$ system.


    \section{ Summary and conclusions }

    Studies of very rare $B$ decays are an excellent way
    to search for New Physics and to constrain New Physics
    model parameters.
    Loop-mediated $B$ decays probe New Physics at high
    mass scales.
    The $B$ factories and the Tevatron experiments have made
    impressive progress which has been matched by our theoretical
    colleagues.
    As of today, there have been no serious challenges to the
    Standard Model from rare $B$ decays.

    The $B$ factories plan to accumulate two to three times more data.
    This will leave many of the clean probes of New Physics
    that I discussed limited by sample statistics, not systematic
    errors.
    The LHC experiments will continue with some but not all
    of these studies.
    Many believe that a super $B$ factory~\cite{superB},
    an ultra-high luminosity
    ($10^{36} \ {\rm cm}^2{\rm s}^{-1}$)
    $e^+e^-$ machine at the $\Upsilon(4S)$ ,would be
    highly desirable, in fact essential, for exploring the 
    flavor sector of any New Physics discovered at the LHC.


   \bigskip 
   \begin{acknowledgments}
   The author would like to thank Joao de Mello and the rest of the
   organizers for hosting a very pleasant conference.
   \end{acknowledgments}
   
   \bigskip 

   \end{document}